\documentclass[conference]{IEEEtran}
\IEEEoverridecommandlockouts
\usepackage{cite}
\usepackage{tabularx,booktabs}
\usepackage{amsmath,amssymb,amsfonts}
\usepackage{algorithmic}
\usepackage{graphicx}
\usepackage{textcomp}
\usepackage{float} 
\newcolumntype{C}{>{\centering\arraybackslash}X}
\usepackage{lipsum}

\def\BibTeX{{\rm B\kern-.05em{\sc i\kern-.025em b}\kern-.08em
    T\kern-.1667em\lower.7ex\hbox{E}\kern-.125emX}}
\begin{document}

\title{Deep Neural Networks with Weighted Averaged Overnight Airflow Features for Sleep Apnea-Hypopnea Severity Classification}

\author{\IEEEauthorblockN{Payongkit Lakhan\IEEEauthorrefmark{1}, Apiwat Ditthapron\IEEEauthorrefmark{2}, Nannapas Banluesombatkul\IEEEauthorrefmark{1} and Theerawit Wilaiprasitporn\IEEEauthorrefmark{1}}
\IEEEauthorblockA{
\IEEEauthorrefmark{1}Bio-inspired Robotics and Neural Engineering Lab, \\School of Information Science and Technology, Vidyasirimedhi Institute of Science \& Technology, Thailand
\\Email: theerawit.w@vistec.ac.th
}
\IEEEauthorblockA{
\IEEEauthorrefmark{2}Computer Department, Worcester Polytechnic Institute, Worcester, MA, USA.}
}

\maketitle

\begin{abstract}
Dramatic raising of Deep Learning (DL) approach and its capability in biomedical applications lead us to explore the advantages of using DL for sleep Apnea-Hypopnea severity classification. To reduce the complexity of clinical diagnosis using Polysomnography (PSG), which is multiple sensing platform, we incorporates our proposed DL scheme into one single Airflow (AF) sensing signal (subset of PSG). Seventeen features have been extracted from AF and then fed into Deep Neural Networks to classify in two studies. First, we proposed a binary classifications which use the cutoff indices at AHI = 5, 15 and 30 events/hour. Second, the multiple Sleep Apnea-Hypopnea Syndrome (SAHS) severity classification was proposed to classify patients into 4 groups including no SAHS, mild SAHS, moderate SAHS, and severe SAHS. For methods evaluation, we used a higher number of patients than related works to accommodate more diversity which includes 520 AF records obtained from the MrOS sleep study (Visit 2) database. We then applied the 10-fold cross-validation technique to get the accuracy, sensitivity and specificity. Moreover, we compared the results from our main classifier with other two approaches which were used in previous researches including the Support Vector Machine (SVM) and the Adaboost-Classification and Regression Trees (AB-CART). From the binary classification, our proposed method provides significantly higher performance than other two approaches with the accuracy of 83.46\%, 85.39\% and 92.69\% in each cutoff, respectively. For the multiclass classification, it also returns a highest accuracy of all approaches with 63.70\%.
\end{abstract}


\section{Introduction}
Obstructive Sleep Apnea-Hypopnea Syndrome (OSAHS) is characterized by repetitive episodes of airflow reduction (hypopnea) or cessation (apnea), which are caused by upper airway collapse during sleep\cite{olson2005}.  Most common symptom of OSAHS is snoring, a sleep disturbance, which results in drowsiness during day time \cite{Javaheri2017}. Furthermore, there are also effects to health qualities such as increasing the risk of Hypertension, Diabetes, Acute Myocardial Infarction, Heart attack, Stroke, Depression, etc.\cite{Javaheri2017}. Polysomnography (PSG) is a clinical measurement technique for the sleep disorder diagnosis \cite{Susheel}. However, multiple physiological signal recordings such as electroencephalogram (EEG), electrocardiogram (ECG), electromyogram (EMG), oxygen saturation of blood (SpO2), leg movement, airflow, cannula flow, respiratory rate and body position are incorporated to PSG \cite{KushidaCA2005}. In general, PSG is performed overnight inside sleep laboratory, either in the hospital or in the clinic \cite{KAKKAR20071057}. Once PSG is recorded, medical doctor with OSAHS expertise need to perform an offline analysis on the whole physiological signals from PSG. Eventually, clinic would report Apnea-Hypopnea Index (AHI) which indicates severity of people with OSAHS\cite{QURESHI2003643}. AHI is catagorized as the following states: no SAHS (Sleep Apnea-Hypopnea Syndrome)  (AHI $<$ 5 events/hr), mild SAHS (5 $\leq$ AHI $<$ 15 events/hr), moderate SAHS (15 $\leq$ AHI $<$ 30 events/hr), and severe SAHS (AHI $\geq$ 30 events/hr) \cite{QURESHI2003643}. Due to complexity and high cost of PSG \cite{JABENNETT,FLEMONS20031543}, one study reported that 90\% of people who had OSAHS were undiagnosed  \cite{Singh2013629}. Thus, simplifying OSAHS diagnosis remains a challenge issue. 

A common approach to solve mentioned issue is reducing the complexity of SAHS (using single physiological signal), cost and analysis time, which are typically required in clinical diagnosis using PSG  \cite{mlynczak2017detecting}. In previous works, researchers did try using single physiological recording from PSG to predict AHI using various computational methods. Single lead ECG, SpO2 from pulse oximeter and airflow (AF) from thermistor were proposed candidates in single recoding for SAHS diagnosis \cite{de2006nasal,magalang2003prediction,penzel2002systematic,nigro2011comparison}. Referred works are based on the same computational strategy which are finding violated periods on the signals and scoring them as apnea-hypopnea related events. The scores are interpreted into AHI eventually. In this study, we aim to develop an automated algorithms to predict SAHS severity by using single time-series, AF which is sensed by the thermistor in front of nose. The comparison of physiological recordings in standard PSG indicated that AF is the most direct measure in breathing obstruction. Moreover, an amplitude of AF will change dramatically during apnea or hypopnea periods [8]\cite{berry2012rules}.

In contrast to aforementioned computational strategy, we implemented our method by using statistical based features together with either classical machine learning approaches (support vector machine, SVM and Adaboost-Classification and Regression Trees, AB-CART) or a modern artificial neural networks (deep neural networks, DNNs). Proposed method begins with statistical extracting features of Apnea and Hypopnea events, and time domain from overnight AF signals. Then, we incorporated features into SVM, AB-CART and DNN for evaluations. Classification tasks had been arranged from simple scenarios which are binary classifications (cutoff indices at AHI = 5, 15 and 30 events/hr). Finally, we performed the same method on multiple classes (no-SAHS, mild-SAHS, moderate-SAHS, and severe-SAHS) afterwards. The experimental studies and performance evaluations were designed according to a previous research on SAHS severity classification using AF signals  \cite{gutierrez2016utility}.

Merits of our works are proposing novel feature extractions from AF signals and performance evaluations on large population. Experimental results of proposed DNNs using proposed features outperformed classical machine learning approaches with the same features. Furthermore, accuracy of proposed DNNs was beyond AB-CART, which was reported to be state-of-the-art for SAHS detection using AF \cite{gutierrez2016utility}.

\section{Methods}
In this section, we first introduce OSAHS datasets from men who parent Osteoporotic Fractures named MrOS \cite{A-dean2016scaling,B-blank2005overview,C-orwoll2005design,D-blackwell2011associations}. Then, we propose statistical based feature extraction from overnight AF signals. The extracted features had fed into SVM and DNN approaches for performance comparison afterwards. 10-fold cross-validation was used to evaluate the performance in both approaches. Binary classifications (three different AHI cutoff indexes) and multi-classes classifications (four severity levels including no-SAHS/control participants) were performed in this study.

\subsection{Datasets}
MrOS sleep data (Visit 2) was used in this study. There were 1,026 men of age 65 years or older participated in standard sleep examinations from six clinical centers. Raw polysomnography (PSG) data in European Data Format (EDF) files with XML annotation files were exported from Compumedics Profusion software. AF signals in PSG were acquired from ProTech Thermistor sensors with 32 Hz sampling rate and high-pass filter at 0.15 Hz cutoff. Each annotation file includes starting and ending times of both Apnea and Hypopnea events. We labeled the severity of SAHS using the AHI variable provided in the datasets. Here, AHI are the numbers of Apnea events in all desaturations and Hypopnea events with 4\% oxygen desaturation per hour\cite{E-chen2015automatic}.

\subsection{Subsampling and Data Preparation}
We did random 520 subjects from the whole 
\begin{table}
\caption {DEMOGRAPHIC DATA FOR THE FOUR-CLASS DIVISION} 
\label{table: Demographic}
\begin{tabularx}{\columnwidth}{@{\extracolsep{2pt}}lCCCCCC@{}}
\cline{1-6}
& no & mild & mod & severe  & All\\ \hline
Subjects & 185 & 190 & 85 & 60 &520 \\
AHI(e/h) & $ 1.82\pm 1.40 $ & $ 8.71\pm 2.97 $ & $ 21.50\pm 3.95$ & $ 41.20 \pm 9.90$ & $12.10\pm13.10$ \\
\hline\hline
\end{tabularx}
\end{table}

datasets for our study. In regard to personal annotation file, we gathered pieces of AF signals during Apnea and Hypopnea events individually. Both AF signals from Apnea and Hypopnea were treated in the same way. In this way, each participant had different numbers and periods of AF samples. Low frequency band is usually a major band in the AF signals \cite{F1alvarez2010spectral}, so low-pass filter with 3 Hz cut-off was applied on all AF samples prior to feature extraction process.

\subsection{Feature Extraction}
After the subsampling and filtering processes, we extracted 17 features from overnight AF samples in the followings:
\begin{itemize}
\item Number of Apnea events.
\item Number of Hypopnea events. 
\item Summation of Apnea and Hypopnea events.
\item Summation of periods (in seconds) from Apnea and Hypopnea events
\item Average of maximum amplitudes from all AF samples.
\item Average of minimum amplitudes from all AF samples.
\item Average of mean amplitudes from all AF samples.
\item Average of standard deviation of amplitudes from all AF samples.
\item Maximum periods from all AF samples.
\item Minimum periods from all AF samples.
\item Mean of periods from all AF samples.
\item Standard deviation of Apnea and Hypopnea periods from all AF samples.
\item Variance of periods from all AF samples.
\item Weighted averaged maximum amplitude from all AF samples
\item Weighted averaged minimum amplitude from all AF samples
\item Weighted averaged mean amplitude from all AF samples
\item Weighted averaged standard deviation of amplitudes from all AF samples
\end{itemize}
\hspace{\parindent} Period of each AF sample is the weight factor in weighted averaging on the last four features.
\begin{figure*}
\centering
\includegraphics[width=0.9\linewidth]{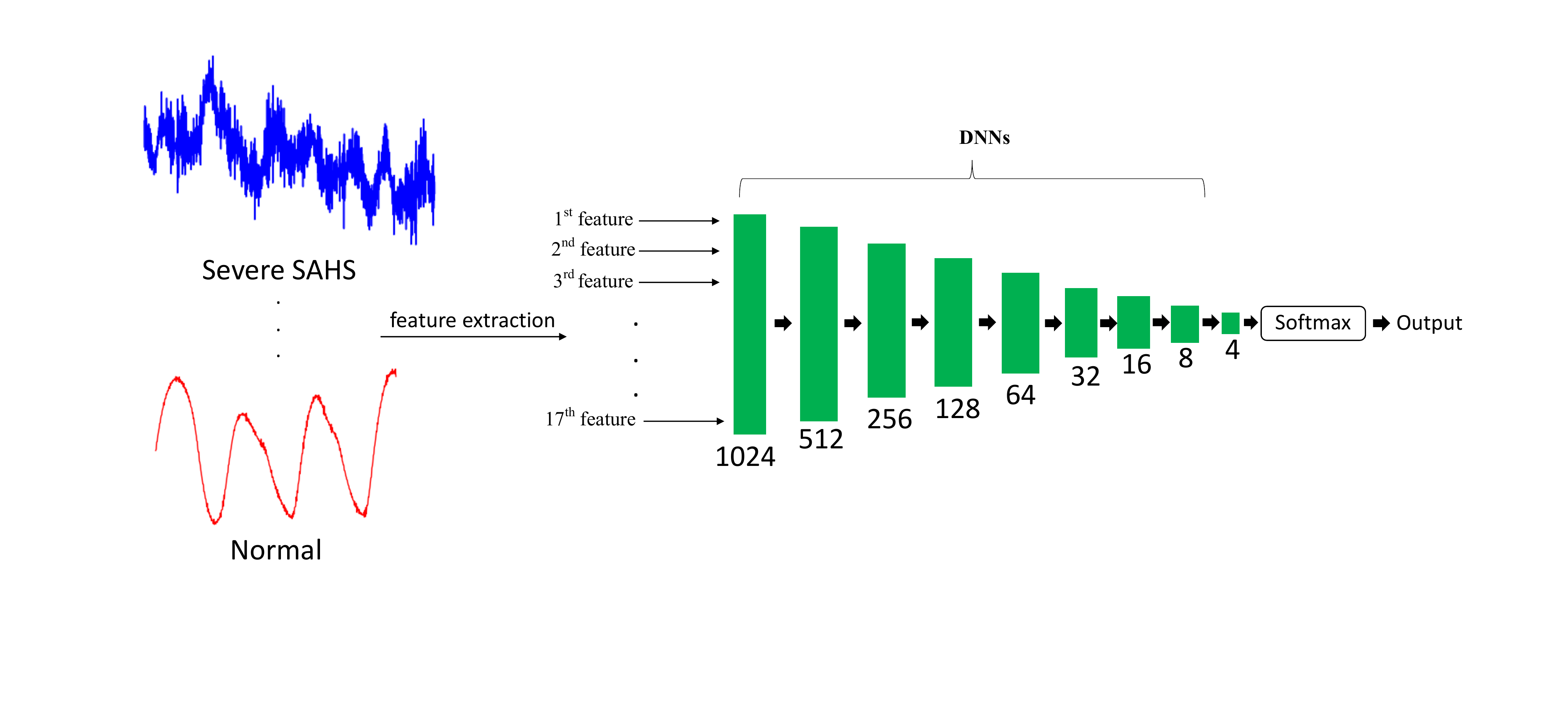}
\caption{Illustration of proposed SAHS severity classifier using seventeen statistical-based features and Deep Neural Networks (DNNs).}
\label{IM1}
\end{figure*}
\subsection{Classification of Sleep Apnea-Hypopnea Syndrome (SAHS) Severity}
There were two main classification tasks in our experiments. First, we aimed to construct three binary classifiers for three AHI cutoff indices, which are clinical standard SAHS cutoff classes (AHI = 5, 15 and 30 events/hr). The number of subjects in this task were shown in Table \ref{table: Demographic}. The second task was to classify subjects in to four standard SAHS classes (normal, mild, moderate and severe). To avoid imbalanced population we did subsampling again in this task, from 520 to be 270 subjects including 70 from normal, mild and moderate subjects and 60 from severe subjects.

Here, we incorporated proposed features into proposed DNNs and classical machine learning (ML) approaches, which are Support Vector Machine (SVM) and Adaboost-Classification and Regression Trees (AB-CART), for comparative classifiers. To validate DNNs, we did split datasets into three subsets which were 80\% for training set, 10\% for validation set and 10\% for testing set. To validate the rest of comparative classifiers, we did use exactly the same datasets as DNNs except validation set. Classical ML does not require validation set or we can simply say that testing set is same as validation set in ML. 10-fold cross-validation had incorporated into these datasets.
\subsubsection{Deep Neural Networks (DNNs)}
As shown in Figure\ref{IM1}, proposed features were fed into the DNNs. DNNs was implemented using Keras API \cite{chollet2015keras} with configuration parameters as follows:
\begin{itemize}
\item A stack of fully-connected neural networks with layer size of 1024, 512, 256, 128, 64, 32, 16, 8, and 4 hidden nodes.
\item Each DNN layer was followed by the Hyperbolic tangent (tanh) activation function.
\item The optimizer was RMSprop with the learning rate of 0.001.
\item The softmax function was applied for classification.
\end{itemize}

\subsubsection{Machine Learning (ML) Approaches}
To present superiority of proposed DNNs over conventional approaches, two classical ML approaches had been used as comparative or baseline classifiers. Conventional SVM with linear kernel and balancing classed weights and AB-CART from scikit-learn API \cite{sklearn_api} had been performed in this study. While SVM is standard baseline classifier, AB-CART had been proposed to use with Airflow (AF)-related sleep Apnea severity in previous study \cite{gutierrez2016utility}.

In the binary classifications, one way repeated measure analysis of variance had been implemented to compare the performance of three classifiers using three metrics: sensitivity, specificity and accuracy. While, confusion matrices had been computed for performance comparison of multiple classes task.
\begin{table*}
\caption {Summary of 10-fold Sensitivity, Specificity and Accuracy of binary classification in each AHI severity cutoff using SVM, AB-CART and DNN classifier. Bold numbers in the table represent the significant highest values in each cutoff.} 
\label{table: summary}
\begin{tabularx}{\textwidth}{@{\extracolsep{1pt}}lCCCCCCCCC@{}}
\hline\hline
& \multicolumn{3}{c}{Sensitivity} & \multicolumn{3}{c}{Specificity}   & \multicolumn{3}{c}{Accuracy}\\
\cline{2-4} \cline{5-7} \cline{8-10} 
&  SVM   & AB-CART & DNN   &  SVM    & AB-CART  & DNN &  SVM    & AB-CART  & DNN    \\ 
\hline
Cutoff 5& $63.79\pm 2.25$ & \boldmath$83.71\pm 1.51$ & $80.47\pm 2.98$&  \boldmath$91.37\pm 5.93$ & $65.24\pm 2.60$& $86.35\pm 1.25$  &$77.89\pm 1.49$ & $77.12\pm 1.42$& \boldmath$83.46\pm 1.08$  \\
Cutoff 15 &\boldmath$90.12\pm 1.44$ &$60.28\pm 2.85$ & $85.56\pm 1.57$  & $51.10\pm 3.36$& \boldmath$90.31\pm 1.33$ & $86.96\pm 3.33$  &  $79.23\pm 1.59$& $78.85\pm 1.15$ &  \boldmath$85.39\pm 1.25$  \\
Cutoff 30 &\boldmath $96.29\pm 0.95$ &$42.95\pm 6.29$& $93.06\pm 0.53$ & $32.29\pm 2.81$  &\boldmath$97.17\pm 0.65$& $90.23\pm 4.23$& $78.07\pm 1.54$ & $90.76\pm 0.94$& \boldmath$92.69\pm 0.55$  \\
\hline\hline
\end{tabularx}
\end{table*}

\begin{table*}
\caption {CONFUSION MATRIX OF 4 CLASSES FROM 10-FOLD CUMULATIVELY} 
\label{table:conf_matrix}
\begin{tabularx}{\textwidth}{@{\extracolsep{2pt}}lCCCCCCCCCCCCCc@{}}
\hline\hline
&  & \multicolumn{4}{c}{SVM} & \multicolumn{4}{c}{AB-CART} & \multicolumn{4}{c}{DNN}&\\
\cline{3-6} \cline{7-10} \cline{11-14}
\multicolumn{2}{c}{Predicted$\rightarrow$} & no & mild & mod & severe & no & mild & mod & severe & no & mild & mod & severe \\ \hline
Actual & no & \textbf{56} & 9 & 4 & 1  &\textbf{41} & 22 & 3 & 4 & \textbf{57} & 12 & 1 & 0\\
& mild  & 20 & \textbf{26} & 17 & 7 &18 & \textbf{26} & 20 & 6& 11 & \textbf{44} & 12 & 3 \\
& mod & 11 & 21 & \textbf{18} & 20  & 12 & 18 & \textbf{24} & 16 & 10 & 17 & \textbf{35} & 8\\
&severe & 4 & 2 & 14 & \textbf{40} &4 & 6 & 17 & \textbf{33}  & 3 & 9 & 12 & \textbf{36}\\
\hline\hline
\end{tabularx}
\end{table*}
\section{Results and Discussion}

After performing binary classification using 3 SAHS severity level cutoffs including AHI = 5, 10 and 15 along with the SVM classifiers, the AB-CART classifiers and our main classifier with or DNN approach, the results are shown in Table \ref{table: summary}. 

For cutoff at AHI = 5, the accuracy of SVM ranges from 76.4\% to 79.38\% (mean $\pm$ standard error, 77.89\% $\pm$ 1.49\%), AB-CART ranges from 75.78\% to 78.54\% (mean $\pm$ standard error, 77.12\% $\pm$ 1.42\%) and our main classifier ranges from 82.38\% to 84.54\% (mean $\pm$ standard error, 83.46\% $\pm$ 1.08\%).

For cutoff at AHI = 15, the accuracy of SVM ranges from 77.64\% to 80.82\% (mean $\pm$ standard deviation error, 79.23\% ± 1.59\%), AB-CART ranged from 77.7\% to 80\% (mean $\pm$ standard error, 78.85\% $\pm$ 1.15\%) and our main classifier ranges from 84.14\% to 86.64\% (mean $w\pm$ standard error, 85.39\% $\pm$ 1.25\%).

For cutoff at AHI = 30, the accuracy of SVM ranges from 76.53\% to 79.61\% (mean $\pm$ standard error, 78.07\% $\pm$ 1.54\%), AB-CART accuracy ranges from 89.82\% to 91.7\% (mean $\pm$ standard error, 90.76\% ± 0.94\%) and our main classifier ranges from 92.14\% to 93.24\% (mean $\pm$ standard error, 92.69\% $\pm$ 0.55\%).

While our DNNs classifiers reached the highest accuracy from all classifiers in every cutoffs and also increased in each of the cutoff, the SVM reached the highest Specificity at AHI = 5 and Sensitivity at AHI = 15 and 30, the AB-CART reached the highest Sensitivity at AHI = 5 and Sensitivity at AHI = 15 and 30. One way repeated measures ANOVA revealed that there were significant difference of mean accuracy among results from three approaches in every cutoffs (AHI=5: \emph{F(2)=2313.822, p$<$0.05}, AHI=15: \emph{F(2)=9.850, p$<$0.05}, AHI=30: \emph{F(2)=50.771, p$<$0.05}). After pairwise comparisons were performed, we found that the accuracies of our DNNs classifiers are significantly higher than others  (p$<$0.05). Consequently, we can conclude that our classifiers are able to sustain the sensitivity and specificity while still maintaining the highest accuracy in all cutoffs.

Additionally, after  the data was balanced in each SAHS level and classified non-linearly into 4 classes, the cumulative confusion matrix of 10 folds are computed as shown in the table for every approaches \ref{table:conf_matrix}. The results from our DNNs classifiers are promising and higher than the others, with the overall accuracy of 63.70\%, while the SVM reached only 51.85\% and the AB-CART reached only 45.93\%. It represents that our DNNs classifier provides a higher diagnostic performance than the other approaches.

\section{Conclusion}
In summary, we proposed statistical based feature extraction from single channel overnight airflow (AF) signals. There are seventeen features in total. Sets of features had fed into proposed DNNs and classical machine learning (ML) approaches, which are Support Vector Machine (SVM) and Adaboost-Classification and Regression Trees (AB-CART), for comparison. Binary and multiple sleep Apnea-Hypopnea severity classifications had been conducted to demonstrate the performance of our proposed features with DNNs which outperformed classical machine learning techniques.

\bibliographystyle{unsrt}
\bibliography{refkim.bib}
\end{document}